\documentclass[prl,twocolumn,showpacs,floats,aps]{revtex4-1}

\usepackage{graphicx}% Include figure files

\usepackage{wasysym}
\usepackage{color}
\usepackage{amsfonts,amssymb}

\usepackage{rotating}

\def\H{{\cal H}}
\def\e{\epsilon}
\def\s{\sigma}

\newcommand{\comment}[1]{}

\begin{document}

\title{Nonequilibrium Zeeman-splitting in quantum transport through nanoscale junctions}

%%% 73.21.La 	Quantum dots
%%  73.63.Rt 	Nanoscale contacts
%%  71.27.+a 	Strongly correlated electron systems; heavy fermions

\author{Sebastian Schmitt}
\author{Frithjof B. Anders}
\affiliation{Lehrstuhl f\"ur Theoretische Physik II, Technische Universit\"at Dortmund,
Otto-Hahn-Str.\ 4, 44221 Dortmund,
Germany}

\date{\today}

\begin{abstract}
We calculate the differential conductance $G(V)$ through a quantum dot in an applied magnetic field. We
use a Keldysh conserving approximation for weakly correlated and the scattering-states numerical
renormalization group for the intermediate and strongly correlated regime out of equilibrium. In the
weakly correlated regime, the Zeeman splitting observable in $G(V)$ strongly depends on the asymmetry of
the device. In contrast, in the strongly correlated regime the position $\Delta_K$ of the Zeeman-split zero-bias
anomaly is almost independent of such asymmetries and of the order of the Zeeman energy $\Delta_0$. We find a
crossover from the purely spin-fluctuation driven Kondo regime at small magnetic fields with $\Delta_K<\Delta_0$ to
a regime at large fields where the contribution of charge fluctuations induces larger splittings with $\Delta_K>\Delta_0$
as it was observed in recent experiments.
\end{abstract}
\maketitle

%%%%%%%%%%%%%%%%%%%%%%%%%%%%%%%%%%%%%%%%%%%%%%%%%%%%%%%%%%%%%%%%%%

Quantum transport through interacting nanoscale junctions has attracted much interest
over the last decade\cite{goldhaberSET98,*MolecularElectronicsBook2005}.
In the Coulomb blockade regime single electron transfer through a quantum dot
coupled to two leads 
is blocked due to the charging energy  $U=e^2/(2C)$
of the device, where $C$ is the capacitance\cite{KastnerSET1992}.
For quantum dots with an odd number of electrons, spin-flip scattering 
opens up a new transport channel due to the Kondo effect\cite{hewson:KondoProblem93}.
The zero-bias conductance 
increases for decreasing temperature and
even reaches the unitary limit value $\frac{2e^2}{h}$ 
characteristic of a fully transparent device\cite{wielNonEqKondo00}.
 
Recent experiments have measured 
the splitting of this zero-bias anomaly (ZBA) of
the differential conductance $G(V)$ in a finite magnetic 
field\cite{quayMagFieldQdot07, jespersenKondoNanowire06, koganMagFieldQdot04, liuMagFieldQdot09}.
While in some experiments
\cite{quayMagFieldQdot07,*jespersenKondoNanowire06} the splitting 
was smaller than the Zeeman splitting 
in accordance with theory,
in others\cite{koganMagFieldQdot04, *liuMagFieldQdot09} a crossover to larger splitting
exceeding the theoretical predictions 
was observed at large magnetic field. 
However, much of the theoretical calculations\cite{loganMagFieldQdot,mooreSpiltMag00,costiQuantumTransport03,hewsonSIAMMagFld06}
have focused on the equilibrium  situation, i.e.\  only the linear response conductance. 
Studies out of equilibrium are usually either perturbative in some 
quantities\cite{hewsonNeqSIAM05,*paaskeNonEqKondoMagField04, *jakobsNeqSIAM10} 
and thus restricted to special parameter values (e.g.\ large or small voltages, small Coulomb 
interaction, etc) or employ methods which do not properly account for the low 
temperature Fermi liquid in equilibrium\cite{meirTransportSIAM93,*balseiroScalingNeqSIAM10}.
Additionally, results obtained for the Kondo-model where
charge fluctuations of the quantum dot are explicitly excluded 
might not be directly applicable to experiments, as such fluctuations 
are inevitably present in actual devices.

In this paper we address the question how a combination of 
charge fluctuations and nonequilibrium effects in strongly correlated systems can indeed
explain the experimentally observed discrepancies. We will pinpoint the crossover
from spin-fluctuation to charge-fluctuation driven transport with increasing bias and fields.
Additionally, we show that the bias dependence of the nonequilibrium spectral function
yields a splitting of the ZBA in the Kondo regime which
is robust against asymmetries 
in contrast to a single-particle picture used in mean-field descriptions.

We employ a Keldysh based conserving
approximation in the weakly
correlated regime\cite{spataruGwMillis2009,SchmittAnders2010} and the
scattering-states numerical renormalization group
(SNRG)\cite{SchmittAnders2010,*AndersSSnrg2008,*AndersNeqGf2008} in
the strongly correlated regime. %  $U/\Gamma>\pi$. 
Both methods are applicable to arbitrary bias $V$ and reduce to the 
corresponding equilibrium theory for $V\to0$.
The SNRG being non-perturbative  covers the full range of interactions
but is numerically much more expensive. 
Its accuracy is determined by discretization errors\cite{BCP08,*AndersSchiller2006}.
Since the SNRG coincides with the Keldysh approach at small $U$
the error remains well controlled as demonstrated in 
Ref.\ \cite{SchmittAnders2010}.
The temperature %in the calculations 
is set much smaller than any characteristic energy 
scale of  the problem and is thus considered as $T\to 0$.

Since the  Kondo temperature $T_K$ is exponentially dependent on $U$\cite{hewson:KondoProblem93}, devices must be  
operated\cite{goldhaberSET98,quayMagFieldQdot07,koganMagFieldQdot04}  
in an intermediate regime with a moderate $U$ to observe the  Kondo effect experimentally.
Consequently, spin and charge fluctuations are not well separated. 
The latter become significant  at  large bias and lead to a modification of the splitting as shown below.
We include the charge fluctuations by considering the single impurity Anderson model (SIAM)
comprising a single orbital on the quantum dot 
coupled to two leads,
\begin{eqnarray}
  \label{eq:siam}
  \H&&=
  \sum_{\sigma, \alpha=L,R} \int \!\!d\e \, (\e -\mu_\alpha) \, c^\dagger_{\e,\sigma\alpha}c^{\phantom{\dagger}}_{\e,\sigma\alpha}
  + \sum_{\sigma = \pm 1}
  E_\sigma   \hat{n}_\sigma
   \\
  && 
  + U\hat{n}_\uparrow\hat{n}_\downarrow
  \nonumber
  + \sum_{\alpha\sigma}\sqrt{\frac{\Gamma_\alpha}{\pi}} \int_{-D}^D\!\! d\e 
  \left\{
    d^\dagger_\sigma c^{\phantom{\dagger}}_{\e\sigma\alpha} +
    c^\dagger_{\e\sigma\alpha} d^{\phantom{\dagger}}_\sigma
  \right\} .
\end{eqnarray}
$d^{\phantom{\dagger}}_{\sigma}$  and $d^\dagger_{\sigma}$ ($c^{\phantom \dagger}_{\e\sigma\alpha}$ and
$c^{ \dagger}_{\e\sigma\alpha}$) are the annihilation and creation operators
for electrons on the quantum dot (in lead $\alpha=L,R$) with spin $\sigma=\pm1$,
$\hat{n}_{\sigma} = d^{\dagger}_{\sigma} d^{\phantom{\dagger}}_{\sigma}$,
and $U$ is the charging energy of the dot. We allow for different coupling  $\Gamma_\alpha$
of  lead  $\alpha$ to the dot, but  fix
the total coupling strength, $\Gamma=\Gamma_L+\Gamma_R$, which is 
used as the unit of energy.
The asymmetry of the coupling is characterized by the ratio  $R=\Gamma_R/\Gamma_L$.
In a finite magnetic field $H$ the single-particle levels of the quantum dot
are shifted by the Zeeman energy $\Delta_0=g\mu_B H$, 
$E_\sigma=E_d+\frac{\sigma}{2}\Delta_0$.
For simplicity, we use a constant density of 
states between $-D$ and $D$ for the noninteracting 
lead electrons 
in the wide-band limit $D\gg U,|E_\sigma|, \Gamma_\alpha$.

The total voltage drop across the junction is given by
the difference in the chemical  potentials of the two leads,
$V=\mu_L-\mu_R$.
In order to model the individual voltage drops at each contact we
employ a  serial resistor model, where the voltage at each contact is inversely
proportional to its coupling strength.
Keeping the dot-levels voltage independent, the chemical potentials in the leads are 
then given by
$\mu_L =\frac{R}{1+R}V$ and $\mu_R=-\frac{1}{1+R}V$. 
Two important limits are incorporated. (i)
For symmetric coupling ($R=1$) half the voltage drops at each side, and
nonequilibrium effects are most pronounced.
(ii) In the tunneling regime   ($R$ or $1/R\to 0$)
only an infinitesimal  current flows through the junction.
The quantum dot is in equilibrium with the stronger coupled lead
and resembles an usual SIAM\cite{hewson:KondoProblem93}, 
where the weaker coupled lead acts only
as a probe similar to a STM tip.

The current $I$ through the quantum dot 
is determined by the nonequilibrium  spectral function 
$\rho_\s(\omega,V)$ \cite{meirLandauerCurrent92}
\begin{eqnarray}
  \label{eq:ss-current}
  I&=&
  \frac{G_0}{e}
 \sum_\s \int \!d\omega  \left[
   f_L(\omega)-f_R(\omega)
  \right]
 \pi\Gamma \rho_\s(\omega,V)
  ,
  \end{eqnarray}
where $G_0 = \frac{e^2}{h}\frac{4R}{(1+R)^2}$, and  
$f_\alpha(\omega)=\frac{1}{1+e^{\beta(\omega-\mu_\alpha)}}$ are
the Fermi functions.
The differential conductance  $G(V)=dI/dV$ 
reveals the Zeeman-splitting of the quantum dot levels in a finite magnetic field.
The peak positions in $G(V)$  depend on 
the energetic level position of the dot, denoted by $\e_{\text{dot}}$, % in the following, 
as well as  on the coupling asymmetry $R$.

In the noninteracting case the dot spectral function remains 
voltage independent. Thus,
the differential conductance at zero temperature
and arbitrary $R$  is
given via the  \textit{equilibrium} spectral function\cite{haugBookTransport96}
\begin{eqnarray}
  \label{eq:gNonInt}
  G_\mathrm{Eq}(V)=
    \comment{
      \pi\Gamma G_0\left[ 
      \frac {\Gamma_L}{\Gamma}\rho_{\mathrm{Eq}}\big(\mu_R\big)
      +\frac{\Gamma_R}{\Gamma}\rho_{\mathrm{Eq}}\big(\mu_L\big)\right]
  }
  \frac{\pi\Gamma G_0}{1+R}\left[ 
    \rho_{\mathrm{Eq}}\big(\mu_R\big)
    +R\,\rho_{\mathrm{Eq}}\big(\mu_L\big)\right]
  ,
\end{eqnarray}
and  maxima in $G_\mathrm{Eq}(V)$ emerge
when one chemical potential coincides with a dot level,  $\mu_\alpha\approx\e_{\text{dot}}$. 
For  $R>1$,  the dominant contribution comes from
the second term in Eq.~(\ref{eq:gNonInt}) and a pronounced peak occurs 
for voltages $V_\mathrm{max}=(1+\frac 1R )\e_{\text{dot}}$.
For symmetric coupling $R=1$, where both terms contribute equally,
a factor of two between peak positions in $G_{\mathrm{Eq}}(V)$ and $\rho_{\mathrm{Eq}}(\omega)$
results,  
$V_\mathrm{max}= 2 \e_{\text{dot}}$,
while in the tunneling regime $R\to\infty$,  $G_{\mathrm{Eq}}(V)\propto \rho_{\mathrm{Eq}}(\omega=V)$.

\begin{figure}[ht]
  \centering
  \includegraphics[width=0.7\columnwidth]{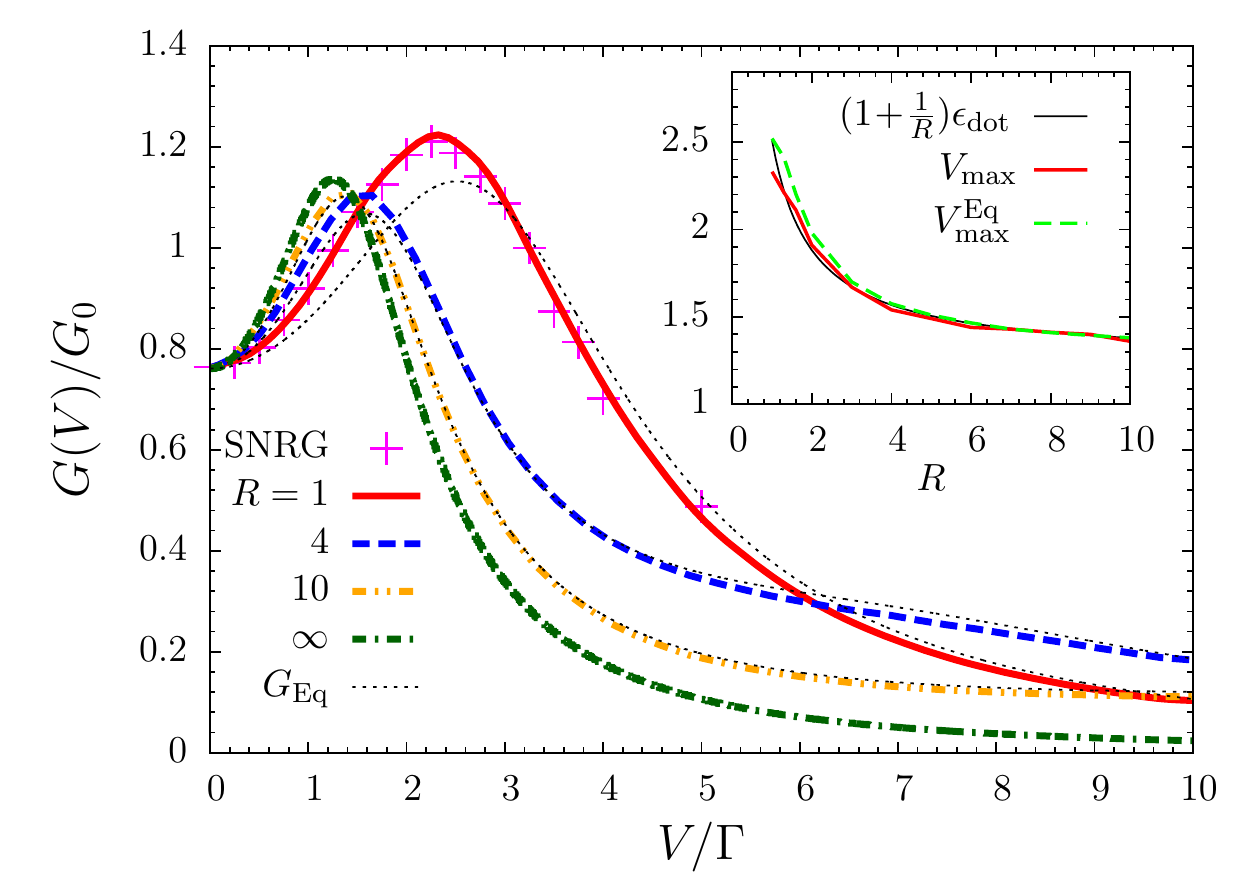}
  \caption{(Color online) 
    Normalized differential conductance $G(V)/G_0$ 
    of a particle-hole symmetric quantum dot  ($E_d=-U/2$) with
    moderate Coulomb interaction $U=\Gamma$ in a finite magnetic field 
    $\Delta_0=2\Gamma$
    and various coupling asymmetry parameters $R$
    as function of bias voltage $V$.
    The curves are calculated in self-consistent 
    second Born approximation within the Keldysh formalism\cite{SchmittAnders2010}. 
    For comparison, the crosses are SNRG result for $R=1$.
    The  estimates $G_\mathrm{Eq}(V)$ 
    are also shown as thin dashed black lines for each $R$.
    The inset shows the  position of maxima as function of $R$.
  }
  \label{fig:G_U1_H2}
\end{figure}

This behavior extends to the weakly correlated regime.
The  differential conductance in a finite magnetic field $\Delta_0=2\Gamma$
is shown in Fig.~\ref{fig:G_U1_H2} for various values of $R$
and a moderate Coulomb interaction $U=-2E_d=\Gamma$.
For comparison, we added the corresponding equilibrium estimation 
$G_{\mathrm{Eq}}(V)$  as thin black dashed lines.
Even though there are pronounced deviations for $R=1$
due to the  $U$-driven voltage dependence in  $\rho_\s(\omega,V)$,
there is an overall agreement between the equilibrium estimate $G_{\mathrm{Eq}}(V)$  and the true 
nonequilibrium conductance $G(V)$.
Both results quickly converge upon increasing $R$ and  
agree almost perfectly already for $R=10$.

\begin{figure*}[t]
  \includegraphics[height=4.7cm]{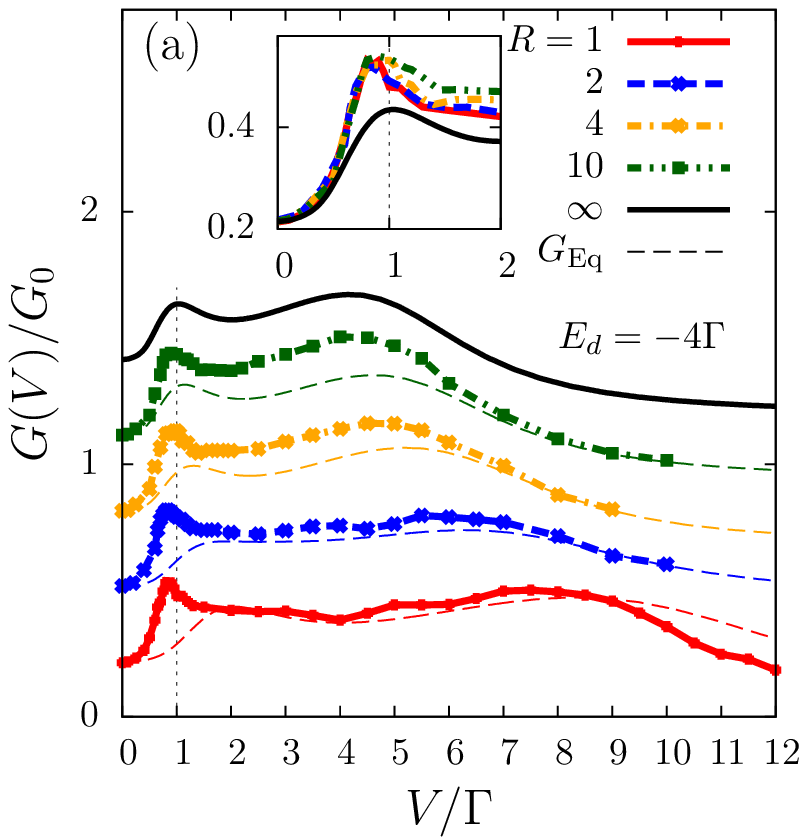}
  \includegraphics[height=4.9cm]{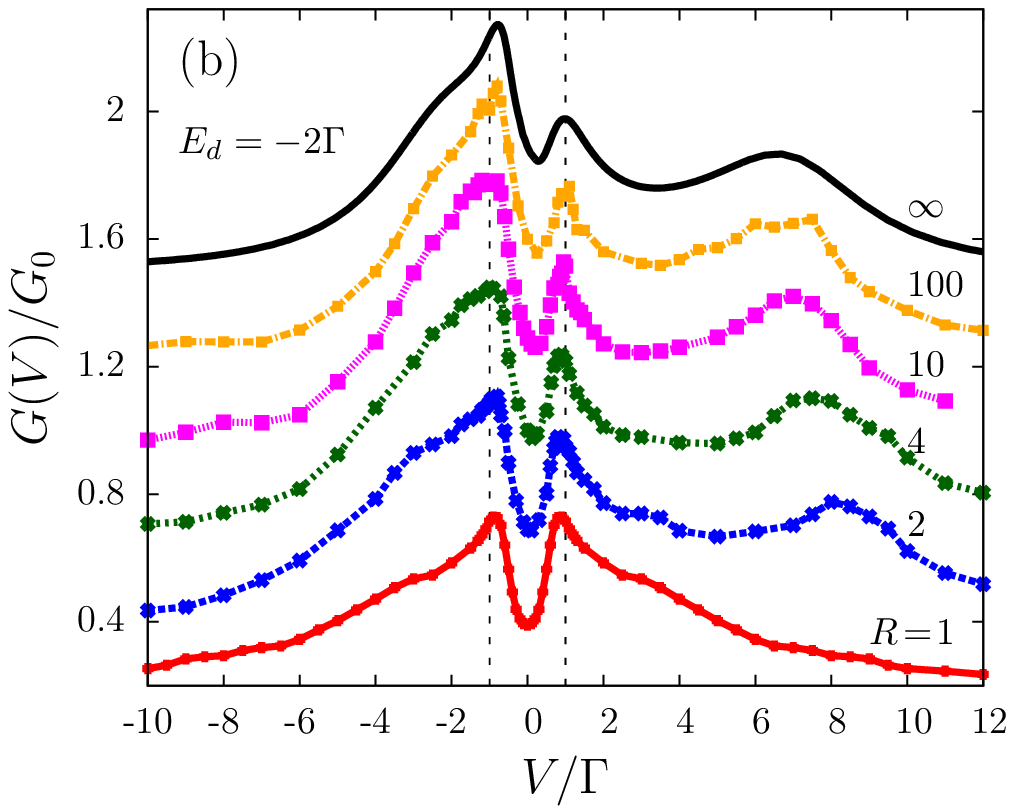}
  \includegraphics[height=4.7cm]{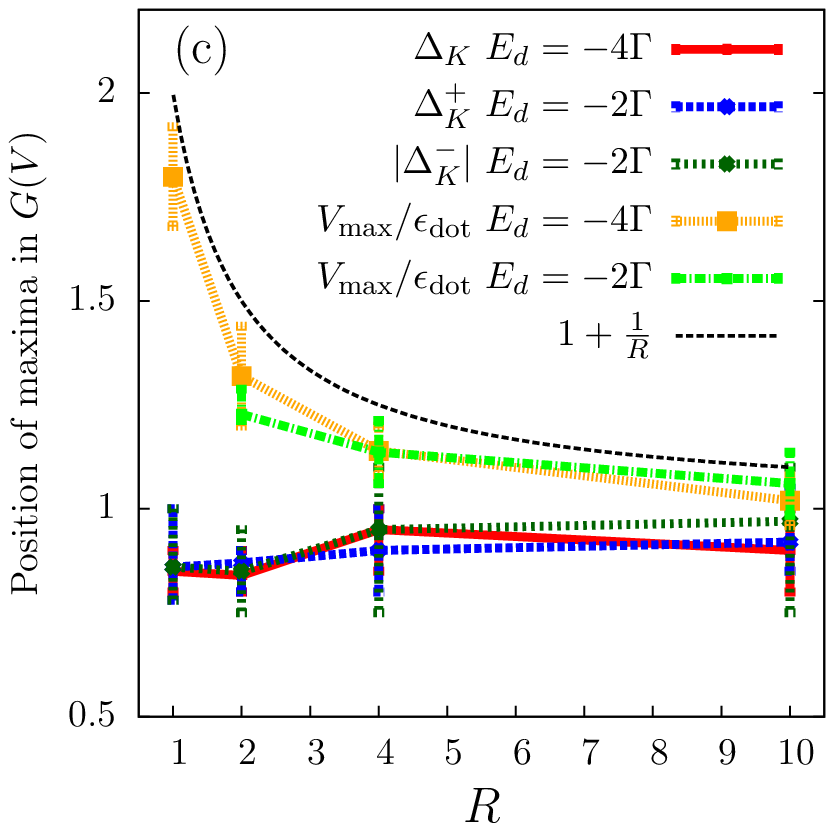}
  \caption{(Color online) Panels (a) and (b): Normalized differential conductance 
    calculated with SNRG for large Coulomb interaction $U=8\Gamma$ 
    in a finite magnetic field, $\Delta_0=\Gamma$, and for various coupling asymmetries.
    (a) $E_d=-4\Gamma$
    (b) $E_d=-2\Gamma$.
    The vertical dashed lines indicate the Zeeman-energy $\Delta_0$ and 
    the different curves  are offset by a constant. 
    The small arrows indicate the position of maxima in $G(V)$, which are plotted in panel (c)
    as function of $R$. $\Delta_K^-$ denotes the position at negative voltages of panel (b). 
}
  \label{fig:didv_U8}
\end{figure*}

Additional to the bare Zeeman splitting  
$\e_{\mathrm{dot}}^{U=0}=\pm\frac{\Delta_0}{2}$
for a noninteracting dot,
the self-energy corrections\cite{SchmittAnders2010}  
caused by the finite  $U$ 
shift  $\e_{\mathrm{dot}}^{U>0}$ to slightly larger values.
For the parameter values of 
Fig.~\ref{fig:G_U1_H2} the peak positions
in $\rho_\mathrm{Eq}(\omega)$ are 
$\e_{\text{dot}}\approx\pm 1.25\Gamma$.
As visible in  the inset,
the maxima in $G(V)$ as well as $G_\mathrm{Eq}(V)$  are located 
approximately at $V_\mathrm{max}\approx (1+\frac 1R) \e_\mathrm{dot}$, and notable 
deviations  occur only near symmetric coupling.
The second contribution from the weaker coupled lead
occurs at $V_\mathrm{max}\approx (R+1) \e_\mathrm{dot}$ and is 
visible for $R=4$ as the small shoulder at $V\approx 6.5\Gamma$.

Fig.\ \ref{fig:didv_U8}(a,b)  shows the 
nonequilibrium conductance of a strongly correlated quantum dot 
in a finite magnetic field 
with  particle-hole symmetric 
and asymmetric 
single-particle energy for various 
%coupling asymmetries 
$R$. 
Only  positive voltages are shown for the particle-hole symmetric dot
of panel (a) due to the symmetry $G(-V)=G(V)$.
The Zeeman energy considerably exceeds the corresponding Kondo temperatures,
$T_K\approx 0.09\Gamma$ for $E_d=-4\Gamma$ and $T_K\approx 0.2\Gamma$
for $E_d=-2\Gamma$.
In both cases, the tunneling regime is perfectly 
approached upon increasing $R$, 
$G(V,R\to\infty)\propto \rho^{\mathrm{NRG}}_{\mathrm{Eq}}(V)$ 
which is added to the plots as black lines.

We clearly can distinguish  two voltage regimes.
At large voltages  $G(V)$  strongly depends on $E_d$  as well as on  the asymmetry $R$. 
The large-voltage maximum is reminiscent of the atomic charge excitation of the dot.
%discussed for the weakly correlated regime above 
Its position 
$V_\mathrm{max}$, as indicated by the arrows in Fig.~\ref{fig:didv_U8}(a,b)
and plotted in Fig.~\ref{fig:didv_U8}(c), is again consistent with  
$V_{\mathrm{max}}\approx (1+\frac1 R)\epsilon_\mathrm{dot}$, where 
$\epsilon_\mathrm{dot}\approx 4.2\Gamma$ ($\epsilon_\mathrm{dot}\approx6.6\Gamma$)
for $E_d=-4\Gamma$ ($E_d=-2\Gamma$) has been obtain
from the equilibrium NRG spectral function $ \rho^{\mathrm{NRG}}_\mathrm{Eq}(\omega)$.

\begin{figure*}[t]
 \includegraphics[width=4.9cm]{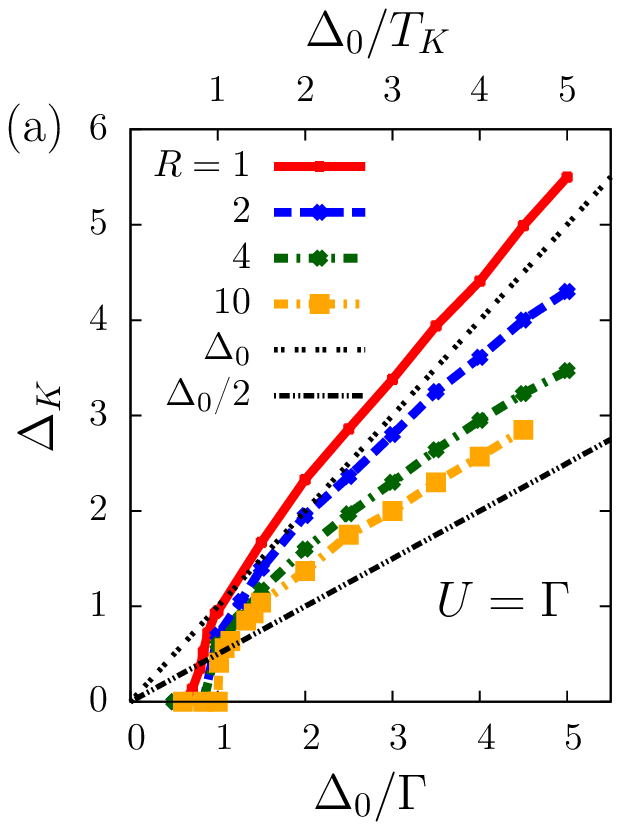}
  \includegraphics[width=4.9cm]{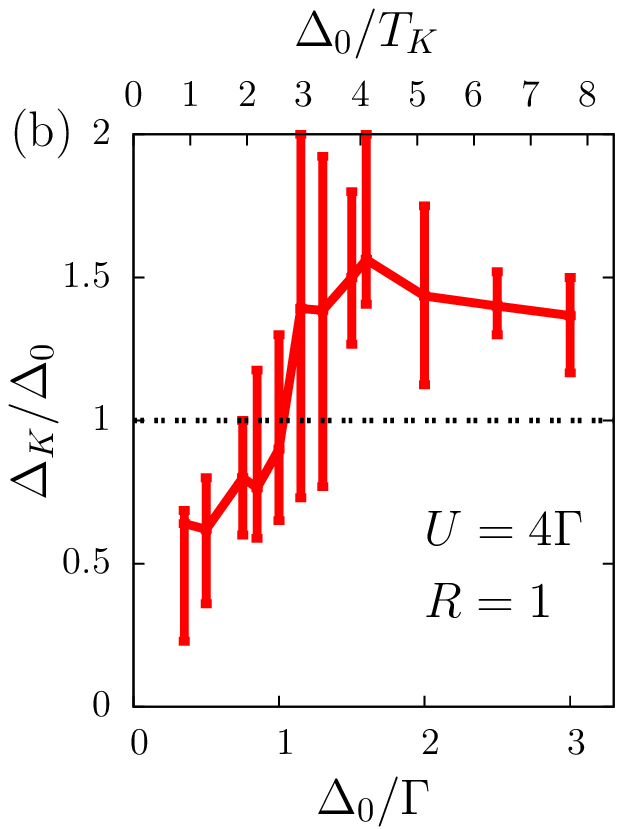}
  \includegraphics[width=4.9cm]{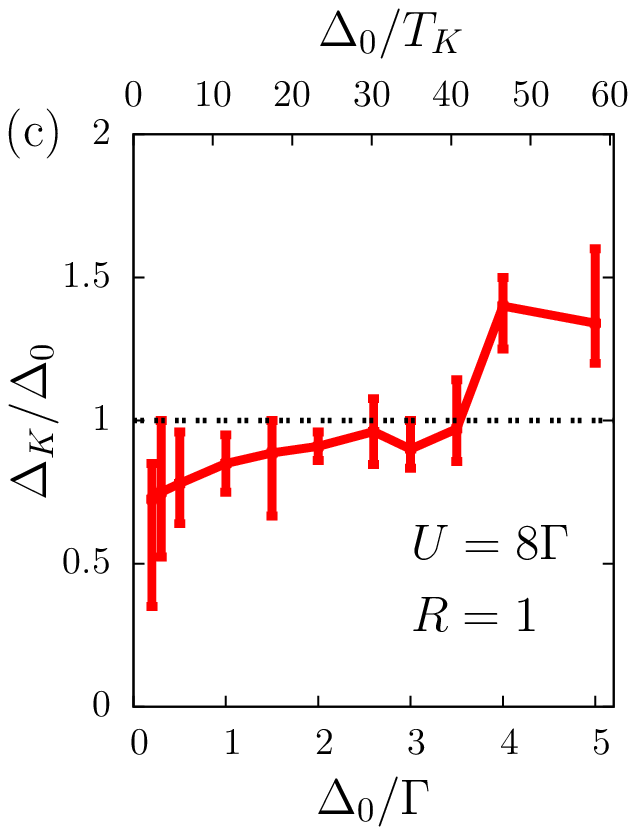}
  \caption{(Color online) Position of the ZBA for particle-hole symmetric quantum dot
    as function of Zeeman energy  for
    Coulomb interaction $U=\Gamma$ (a),  $4\Gamma$ (b), and  $8\Gamma$ (c).
    Panel (a) directly displays $\Delta_K$ calculated with the second Born approximation for various $R$, while
    panels (b) and (c) show the normalized position $\Delta_K/\Delta_0$
    for a symmetric junction $R=1$ obtained within the SNRG. 
    The inaccuracies 
    due to the numerical differentiation of $I(V)$ are indicated by the error bars in panels (b) and (c).
    The  scale at the bottom measures $\Delta_0$ in units of the
    coupling $\Gamma$, while the top scale is in units 
    of the Kondo temperature $T_K$. 
  }
  \label{fig:peakPos}
\end{figure*}

In contrast, the position of the Zeeman-split ZBA $\Delta_K$ 
at low voltage (also indicated by arrows) appears to be almost \textit{independent} of particle-hole 
and coupling asymmetry as clearly seen in 
Fig.\ \ref{fig:didv_U8}(a,b). 
$\Delta_K$ only slightly increases with increasing $R$ [see Fig.~\ref{fig:didv_U8}(c)]
and is essentially 
given by the Zeeman-energy $\Delta_K\approx \Delta_0$.
In the linear response regime\cite{loganMagFieldQdot,hewsonSIAMMagFld06,mooreSpiltMag00,*costiQuantumTransport03}
the spectral function  is approximated to be voltage independent: $\rho(\omega,V)\approx \rho_\mathrm{Eq}(\omega)$.
This  would imply that the maximum
of the Zeeman-split ZBA in $G(V)$  should occur at $\Delta_K\approx(1+\frac{1}{R}) \omega_{\mathrm{Kondo}}$,
where $\omega_{\mathrm{Kondo}}$ is the  position of the Zeeman-split Kondo resonance
in $\rho_\mathrm{Eq}(\omega)$.
As a hallmark of the Kondo effect, the splitting of the Kondo resonance in 
$\rho_\mathrm{Eq}(\omega)$
is strongly enhanced and approaches \textit{twice} the value of a noninteracting level for large fields,
$\omega_{\mathrm{Kondo}}\approx \Delta_0$\cite{loganMagFieldQdot,*costiQuantumTransport03,hewsonSIAMMagFld06}.
The position of the Zeeman-split ZBA would thus evolve with $R$
from $\Delta_K\approx\Delta_0$ in the tunneling regime to   
$\Delta_K\approx 2\Delta_0$ for a symmetric junction. 
%and a voltage-independent spectrum.
Therefore, this simple picture which partially extents to 
the high-voltage behavior completely fails for the Kondo-correlated ZBA!

The ZBA arises from magnetic spin-flip
scattering\cite{glazmanLowTempTrans05}, 
where the spin of the electron on and the spin of the electron scattered across the dot 
contributing to the current are both flipped. 
In a finite external magnetic field such a spin-flip 
is associated with a finite energy cost $\Delta_0$,
which must be provided by the bias voltage for the 
transport channel to open, $V\gtrsim \Delta_0$.
Hence, the asymmetry governs only the
magnitude of the peak (and might lead to a merge with a charge excitation peak 
as in Fig.~\ref{fig:didv_U8}(b) for negative voltages) but it does not alter 
its position  as this is determined in leading order only by the difference in  energy
between the initial and final state of the dot. 

So the well-known and established picture of spin-flip scattering as the driving force behind
the Kondo-effect provides a clear understanding of the asymmetry-independence.
However, the insensitivity of the ZBA to asymmetries
observed here as well as in 
experiments\cite{quayMagFieldQdot07,jespersenKondoNanowire06,liuMagFieldQdot09,koganMagFieldQdot04}
is  highly nontrivial, if viewed in the light of the Meir-Wingreen formula (\ref{eq:ss-current}). 
Obviously, a simple translation of  maxima in the equilibrium spectra
$\rho_\mathrm{Eq}(\omega)$,  to maxima
in the \textit{nonequilibrium} differential conductance $G(V)$ 
fails, and a full understanding of the bias dependence of 
many-body correlations  is required. 
The voltage dependent renormalization of the transmission is 
essential for a correct description 
of the conductance as already pointed out\cite{hewsonSIAMMagFld06}.
In the Kondo regime, our non-perturbative 
many-body calculation out of equilibrium has revealed
that 
the redistribution of spectral weight
in $\rho(\omega,V)$\cite{SchmittAnders2010}
with bias occurs in such a way that the 
maximum in $G(V)$ is always found at $\Delta_K\approx \Delta_0$,  
irrespective of $R$ and $E_d$.

Fig.\ \ref{fig:peakPos}(a) summarizes
the results for $\Delta_K$ as function of the applied magnetic field
for  small interaction $U/\Gamma=1$ as used in Fig.\ \ref{fig:G_U1_H2}.
The finite width of the ZBA determines the threshold field
above which the splitting can be observed. 
Here,  Kondo correlations are small, and the slope
of $\Delta_K$ approaches $\frac 12 (1+\frac 1 R)$
at large fields. The constant offset indicates the Hartree shift due to a finite $U$.

Fig.\ \ref{fig:peakPos}(b,c) displays the normalized peak position $\Delta_K/\Delta_0$ 
of the ZBA  for a symmetric junction with $U/\Gamma=4$, $8$.
In both cases, the zero-field extrapolation is consistent with
the asymptotic value of the equilibrium Kondo regime\cite{loganMagFieldQdot}, 
$\lim_{\Delta_0\to 0}\Delta_K/\Delta_0=\frac 23$.
As in equilibrium\cite{costiQuantumTransport03}, $\Delta_K/\Delta_0$ increases with 
field,
and in the strongly correlated regime [panel (c)] this increase in rather slow 
due to the presence of typical Kondo-logarithms.
In contrast to the equilibrium Kondo regime, however,  a crossover to a 
non-universal behavior $\Delta_K/\Delta_0 > 1$
occurs at fields of the order of the charge scale, $\Delta_0\sim\frac U2 =4\Gamma\approx 45T_K$.
For  $U=4\Gamma$ [panel(b)], the charge fluctuations dominate
the physics much earlier, and the crossover is observed already
at $\Delta_0\sim\Gamma\approx 3T_K$.
We attribute this non-universal Zeeman splitting at large field
to the nonequilibrium charge fluctuations.  

We can distinguish two different regimes for strong correlations $U> \pi\Gamma$:
(i) At fields and voltages below the charge excitation scale, Kondo correlations dominate 
leading to a renormalization of $\Delta_K< \Delta_0$, (ii) at high fields 
the Zeeman split Kondo peaks merge with the Hubbard side bands of $\rho(\omega,V)$
and the Kondo effect is destroyed.
The physics is mainly driven by charge fluctuations accessible at large voltages
and  $\Delta_K > \Delta_0$
similar to the weakly correlated regime. 

We have indeed reproduced the qualitative findings of experiments\cite{koganMagFieldQdot04,liuMagFieldQdot09} and
have identified the  non-universal splitting as 
consequence of charge fluctuations.
However, for $U/\Gamma\approx 8$ as estimated from experimental
setups, the crossover in our calculations 
occurs at $\Delta_0/T_K\approx 40$ [panel (c)]. 
This is an order of magnitude larger than the values reported in experiments
where the crossover typically occurs at $\Delta_0/T_K\approx 3-7$. 
We can reproduce 
such values only with a smaller Coulomb interaction  $U/\Gamma=4$.
This might be hint for an additional transport
mechanism present in experiments, which is not accounted for
in the simple SIAM  investigated here.

In summary, we have investigated the
% magnetic field
Zeeman splitting of the ZBA for weak,
intermediate and strongly correlated nanodevices under true nonequilibrium conditions.
In weakly correlated devices or generally at large bias the transport is
driven by charge fluctuations, and the splitting strongly depends on the junction asymmetry.
In contrast, for the Kondo-correlated ZBA present at large $U/\Gamma$
our approach predicts an almost asymmetry-independent splitting,
which is in accord with experiments and can be understood in a simple spin-flip scattering picture.
Our theory explains the experimentally observed 
crossover\cite{koganMagFieldQdot04,liuMagFieldQdot09} from $\Delta_K<\Delta_0$ to $\Delta_K>\Delta_0$ from spin to charge fluctuation driven
transport, but also reveals shortcomings of an
oversimplified description using only a single level due to the inaccurate scale for this crossover.

%\section{Acknowledgments}
We are grateful to  F.~G\"uttge, J.~Han,  A.C.~Hewson J. Paaske, 
A.~Schiller and   P.~Schmitteckert  for helpful discussions.
We acknowledge financial support from the Deutsche Forschungsgemeinschaft under AN  275/6-2
and supercomputer support by the NIC, FZ J\"ulich under project No.\ HHB00. 

%\bibliography{references}

%%%%%%%%%%%%%%%%%%%%%%%%%%%%%%%%%%%%%%%%%%%%%%%%%%%%%%%%%%%%%%%%%%%%%%%%%%%%%%%%%%%%%%%%%%%%
%
\end{document}